\documentclass[aps,superscriptaddress,preprint]{revtex4-1}
\usepackage{amsmath}
\usepackage{graphicx}
\usepackage[version=3]{mhchem}
\usepackage{pdfpages}
 \usepackage{amssymb}
 \setcitestyle{super}

\begin{document}

\title{Efficient Direct Solar-to-Hydrogen Conversion by \textit{In Situ} Interface Transformation of a Tandem Structure}

\author{Matthias M. May}
\affiliation{Helmholtz-Zentrum Berlin f\"ur Materialien und Energie GmbH, Institute for Solar Fuels, Hahn-Meitner-Platz~1, D-14109 Berlin, Germany}
\affiliation{Technische Universit\"at Ilmenau, Department of Physcis, Gustav-Kirchhoff-Str.~5, D-98693 Ilmenau, Germany}
\affiliation{Humboldt-Universit\"at zu Berlin, Department of Physcis, Newtonstr.~15, D-12489 Berlin, Germany}
\author{Hans-Joachim Lewerenz}
\affiliation{California Institute of Technology, Joint Center for Artificial Photosynthesis, 1200 East California Boulevard, CA 91125 Pasadena, USA}
\author{David Lackner}
\affiliation{Fraunhofer Institute for Solar Energy Systems ISE, Heidenhofstr.~2, D-79110 Freiburg, Germany}
\author{Frank Dimroth}
\affiliation{Fraunhofer Institute for Solar Energy Systems ISE, Heidenhofstr.~2, D-79110 Freiburg, Germany}
\author{Thomas Hannappel}
\affiliation{Technische Universit\"at Ilmenau, Department of Physcis, Gustav-Kirchhoff-Str.~5, D-98693 Ilmenau, Germany}


\begin{abstract}

Photosynthesis is nature's route to convert intermittent solar irradiation into storable energy, while its use for an industrial energy supply is impaired by low efficiency. Artificial photosynthesis provides a promising alternative for efficient robust carbon-neutral renewable energy generation. The approach of direct hydrogen generation by photoelectrochemical water splitting utilises customised tandem absorber structures to mimic the Z-scheme of natural photosynthesis. Here, a combined chemical surface transformation of a tandem structure and catalyst deposition at ambient temperature yields photocurrents approaching the theoretical limit of the absorber and results in a solar-to-hydrogen efficiency of 14\%. The potentiostatically assisted photoelectrode efficiency is 17\%. Present benchmarks for integrated systems are clearly exceeded. Details of the \textit{in situ} interface transformation, the electronic improvement and chemical passivation are presented. The surface functionalisation procedure is widely applicable and can be precisely controlled, allowing further developments of high-efficiency robust hydrogen generators.

\end{abstract}

\maketitle

Hydrogen provides the highest energy density among the common fuels\cite{Parkinson_PEC_global_energy_2013}. In the search for sustainable, low-carbon replacements for fossil fuels, solar water splitting is receiving exceptional attention, with \ce{H2} as the crucial ingredient of an anthropogenic carbon cycle or a completely carbon-free energy economy\cite{Parkinson_PEC_global_energy_2013, Olah_anthropogenic_carbon_cycle_2011}. Solar energy is stored in chemical bonds by splitting H$_2$O photoelectrochemically into H$_2$ and O$_2$. The associated thermodynamic potential difference of 1.23\,V must be exceeded by the free energy of the charge carriers transferred to the electrolyte\cite{Fujishima_photolysis_1972}. The required, minimum photovoltage for the photolysis of water is $\gtrsim1.6$\,V, including catalyst overpotentials, and depends on the chosen photocurrent density\cite{Gerischer_heterogeneous_elchemical_systems_1980}. Since excess energy is dissipated as heat, the chemical energy stored in a single H$_2$ molecule is $2\times1.23$\,eV. Overcoming the voltage threshold and utilising a constant energy per reduced hydronium ion requires photocurrent maximisation, if the photovoltage suffices to drive the reaction. These boundary conditions make tandem photovoltaic devices, where semiconductor absorber layers with different energy gaps are combined for enhanced exploitation of the solar spectrum, superior candidates for photoelectrochemical water splitting. As a highly efficient, inorganic analogue to the Z-scheme of natural photosynthesis\cite{Bard_z-scheme_1979}, dual tandem structures exhibit increased photovoltages while simultaneously exploiting the sunlight efficiently, thus permitting high photocurrents. In contrast, photovoltaic power generation is less restricted as the utilisable output power is defined via the current-voltage product. Stacking multiple absorbers can increase both photovoltage and delivered power at the expense of the achievable current\cite{Shockley_detailed_balance_1961, PEC_book_chapter_2013, Wuerfel_physics_of_solar_cells_2009}. Despite employing relatively high-cost, but high-efficiency structures, solar \ce{H2} is predicted to become competitive at solar-to-hydrogen (STH) efficiencies of 15\% and beyond\citep{Pinaud_hydrogen_production_facilities_2013}.

Two principle approaches for solar water splitting have been advocated: The non-monolithic approach, where photovoltaic light harvesting and electrolytic water splitting are spatially separated, or monolithically integrated devices, which are fully immersed into the electrolytes. The former one avoids issues associated with the semiconductor--electrolyte contact, such as the challenge to employ heterogeneous catalysts with low light absorption\cite{Seger_tandem_designs_2014} and photocorrosion, but necessitates a second technology line\cite{Bard_artificial_photosynthesis_1995}. Whereas with the former approach, STH efficiencies of up to 18\% have been achieved\cite{Licht_2001, Luo_Perovskite_water_splitting_2014}, the benchmark for monolithic water splitting is at 12.4\% STH for 11 suns illumination, and achieving simultaneously efficiency and stability remains an issue \cite{Khaselev1998}.

For the integrated, direct approach pursued here, the photolysis cell design is more demanding, but is alleviated by a higher potential for cost reduction of solar \ce{H2} generation\cite{Pinaud_hydrogen_production_facilities_2013, Zhai_net_energy_balance_water_splitting_2013}. An extensive overview of existing systems and their performance can be found in the literature, though sometimes no clear distinction between monolithic and non-monolithic systems is made there\cite{Ager_review_STH-efficiencies_2015}.

Selection of the tandem absorber energy gaps for an efficient use of the solar spectrum and simultaneous supply of the necessary photovoltage for water photolysis yields optimum energy gap combinations in the range of 0.8-1.2\,eV and 1.5-1.9\,eV for the bottom and top cells, respectively\cite{PEC_book_chapter_2013, Hu_optimal_band_gaps_PEC_2013}. For III-V semiconductors, their adjustable optoelectronic properties, the high control of doping levels, the formation of tunnel junctions and abrupt interfaces enable such an adaptation\cite{Vurgaftman_band_parameters_III-V_2001, Dimroth_metamorphic_cell_2001, Sagol_basic_concepts_multijunction_2007}. Application in photoelectrochemical devices necessitates in addition a careful conditioning of the interfaces of the absorber with the electrolyte and with the electrocatalyst, also considering molecular details of the surface chemistry\cite{May_GaP_H2O_2013, May_InP_H2O_2014}. The interface modification has to simultaneously provide corrosion protection, high optoelectronic quality, sufficient optical transparency as well as a suitable mechanical and electronic coupling to the catalyst\cite{PEC_book_chapter_2013, Lewerenz_EESci_2010, Wood_surface_chemistry_water_InP_GaP_2014}. \textit{In situ} (photo)electrochemical functionalisation is a low-temperature, ambient pressure wet processing method. Similar to galvanic processing, it is widely applicable and, in principle, industrially scalable. It has already been shown that wet semiconductor processing can provide high-quality interfaces combined with high stability\cite{PEC_book_chapter_2013, Lewerenz_EESci_2010}.

We present here an \textit{in situ} surface functionalisation routine developed for a III-V photovoltaic tandem absorber, that enables an STH of 14\% for unbiased, direct solar water splitting. Electronic and chemical passivation are achieved via a transformation of the surface AlInP layer towards oxides and phosphates/phosphites, which allow efficient coupling to the Rh \ce{H2} evolution electrocatalyst. A further reduction of interfacial charge-carrier recombination would unlock the full potential of the device reaching efficiencies beyond 17\%, which is predicted to become interesting for commercial prototypes\cite{Pinaud_hydrogen_production_facilities_2013}.

\section*{Results}
\subsection*{Customised tandem absorber}

\begin{figure}
\includegraphics{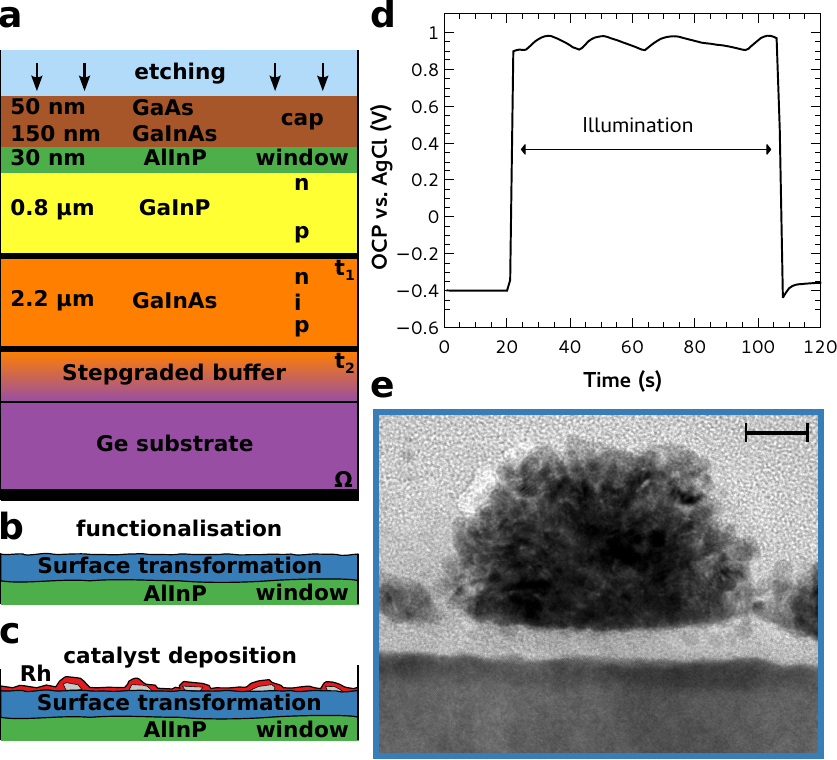} 
\caption{\textbf{Interfacial functionalisation steps.} The tandem consists of a GaInP \textit{n-p} top cell and a GaInAs \textit{n-i-p} bottom cell. (a) Stacking and thicknesses; t$_1$, t$_2$, and $\Omega$ denote tunnel junctions and ohmic back contact, respectively; vertical arrows indicate selective chemical etching of the capping layer. (b) Chemical and photoelectrochemical surface transformation of the AlInP window layer in an aqueous solution of RhCl. (c) Electrochemical deposition of a seed layer of Rh electrocatalysts (grey) and photoelectrodeposition of a continuous electrocatalyst film (red) (see text). (d) Oscillation of the open-circuit potential (OCP) upon illumination during functionalisation. (e) Transmission electron microscope image of a cross-section of the surface after Rh catalyst deposition, the scale bar indicates 10\,nm.}
\label{fig:fig1_processing}
\end{figure}

\begin{figure}
\includegraphics[width=\linewidth]{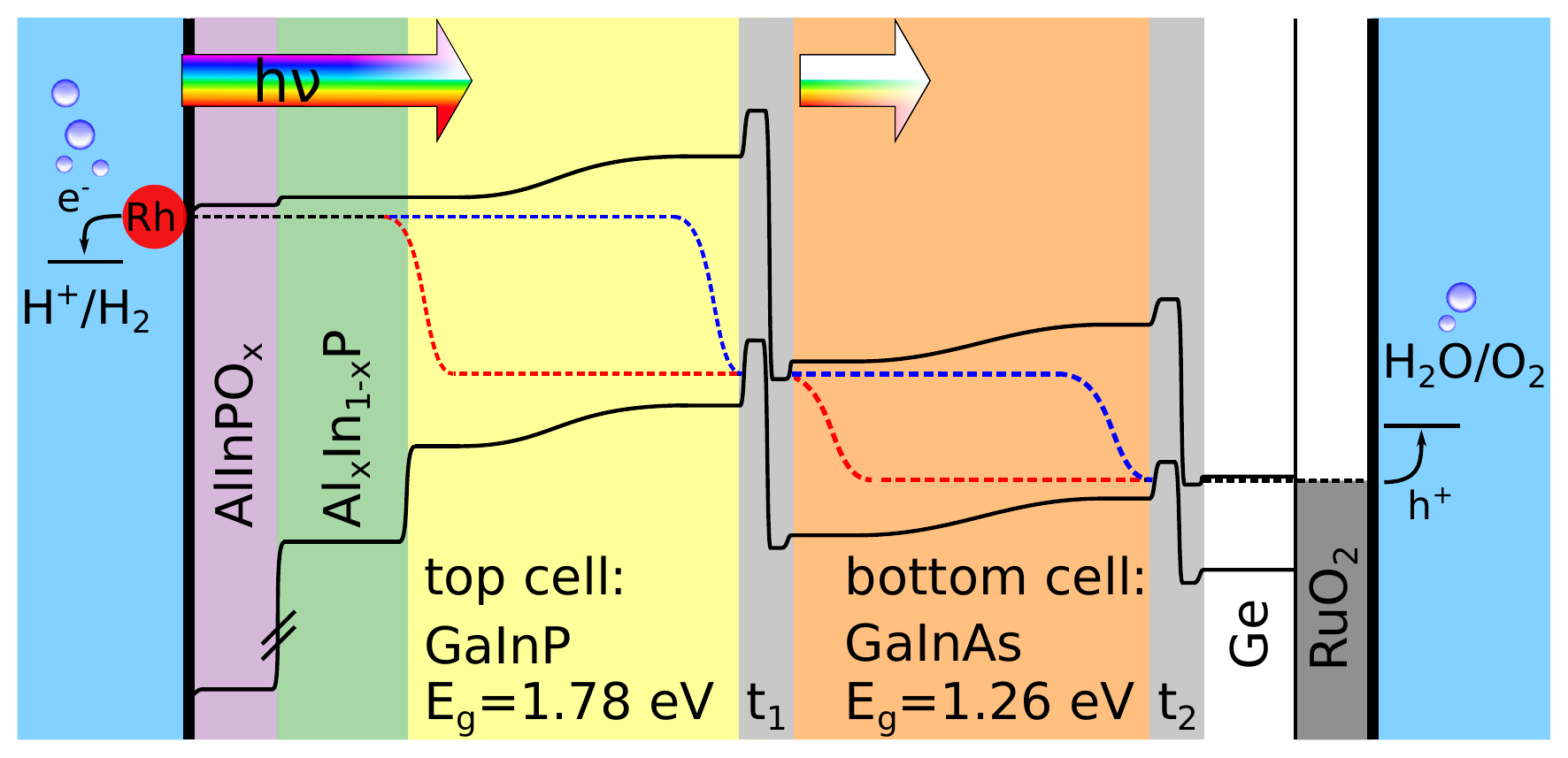}
\caption{\textbf{Energy schematic of the tandem layer structure under illumination.} The ohmic back-contact is connected to a sputtered RuO$_2$ counter-electrode, where \ce{O2} evolution occurs. \ce{H2} evolution takes place at the interface of the Rh electrocatalyst on top of the functionalised layer with the electrolyte. Ohmic contacts between subcells are made by tunnel junctions, where switching between majority-charge carrier types occurs. Black, dashed lines represent the Fermi level, blue (red) the Quasi-Fermi levels of electrons (holes) and arrows indicate the impinging light.}
\label{fig:device_structure}
\end{figure}

A two-junction tandem absorber structure, grown epitaxially on a Ge substrate by metal-organic vapour phase epitaxy\cite{Dimroth_metamorphic_cell_2001}, serves as photovoltaic core element. The absorber consists of a GaInP \textit{n-p} top cell with an energy gap of $E_g=1.78$\,eV and of an \textit{n-i-p} GaInAs bottom cell with $E_g=1.26$\,eV (Fig.~\ref{fig:fig1_processing}a). The structure also contains cap layers and an AlInP window layer for electron collection. In the photovoltaic mode and with anti-reflection coating, a short-circuit current density of up to 15\,mAcm$^{-2}$ and an open-circuit voltage of 2.1\,V is expected at AM 1.5G illumination\cite{Dimroth_metamorphic_cell_2001}. The energy band relations are depicted in Fig.~\ref{fig:device_structure}. They show the energetic situation near the maximum power point of the device where, for high fill factors, the photovoltage is close to the open-circuit voltage, but the photocurrent is still near its maximum (short circuit) value. The photovoltage generated is represented by the difference of the quasi Fermi levels of electrons (blue curve in Fig.~\ref{fig:device_structure}) and holes (red curve), yielding the achievable free energy of the system near the maximum power point. The structure operates as photocathode with a dark anode using a \ce{RuO2} catalyst for \ce{O2} evolution.

\subsection*{Interface functionalisation}

\begin{figure}[h]
\includegraphics[width=.5\textwidth]{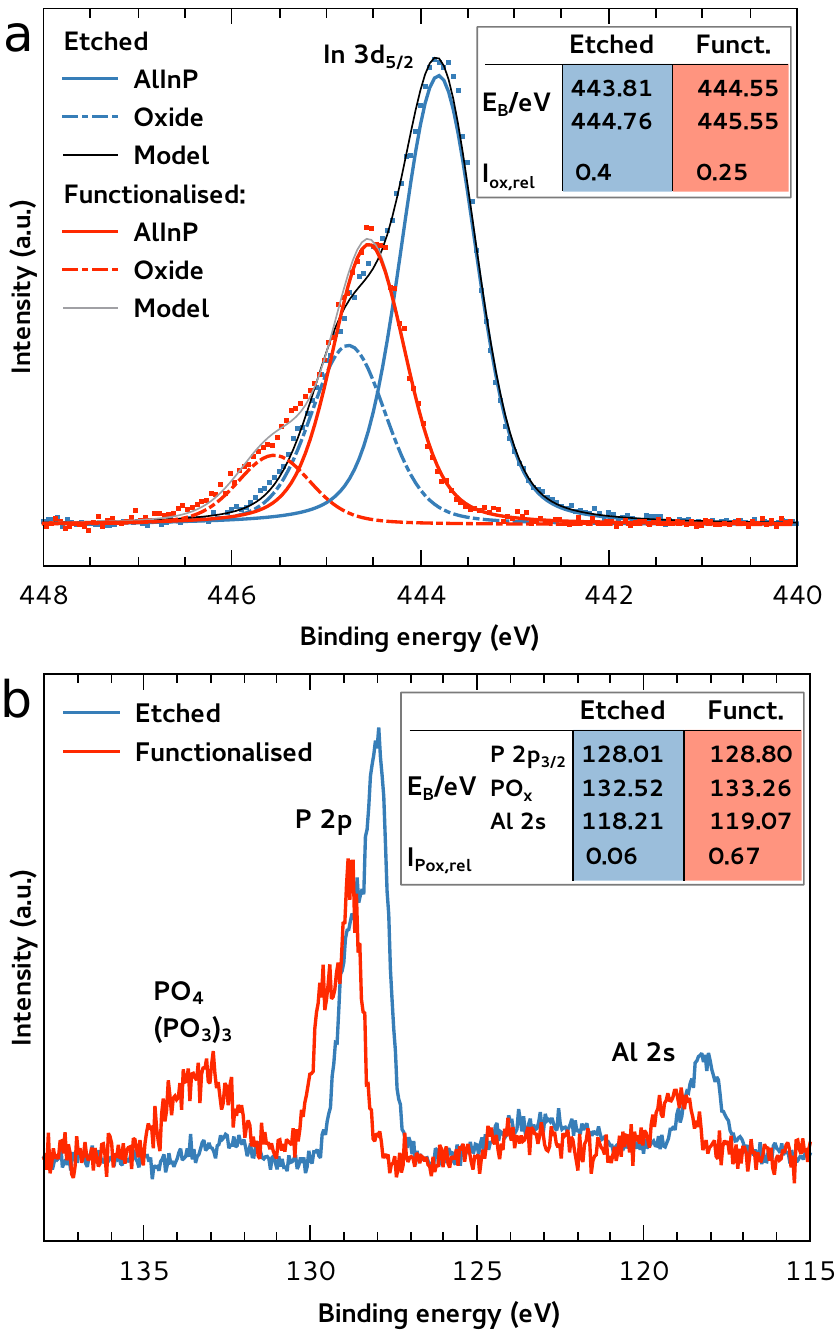} 
\caption{\textbf{Surface analysis after functionalisation by XPS.} (a) Spectrum of the In 3d 5/2 core level after chemical etching of the capping layer, before (blue curve) and after (red curve) complete functionalisation revealing a decrease of In content and composition (see text). (b) P 2p and Al 2s core levels unveiling a transformation of P in AlInP towards phosphite/phosphate species.}
\label{fig:XPS}
\end{figure}

The fundamental surface processing steps are summarised in Fig.~\ref{fig:fig1_processing}a-c. A protective cap layer was removed by selective etching\cite{Clawson_III-V_etching_2001} in a solution of NH$_4$OH:H$_2$O$_2$:H$_2$O, stopping at the n$^+$-doped \ce{Al_{0.35}In_{0.65}P} window layer. A suitable surface for the subsequent functionalisation was prepared by chemical oxidation in O$_2$-saturated water, followed by a drying process in a flow of oxygen, resulting in a smooth, oxidised surface. The surface composition was analysed by X-ray photoelectron spectroscopy (XPS), revealing a mixed In-/Al-oxide layer on top of the AlInP window (see Fig.~\ref{fig:XPS}). Such oxide layers typically form a barrier for holes, reducing charge carrier recombination\cite{Lewerenz_EESci_2010, PEC_book_chapter_2013}.

Combined chemical and photoelectrochemical surface conditioning is essential to increase stability and charge-transfer efficiency, as the stability of a chemically formed, pure group III element (In, Al) oxide layer, which is the outcome of the etching process (Figs.~\ref{fig:fig1_processing}a and \ref{fig:XPS}), can be rather limited under operating conditions\cite{Khaselev_stability_1998}. We developed an advanced, but straightforward procedure that allows for simultaneous surface functionalisation and catalyst deposition in a single  solution in a two-step approach (Figs.~\ref{fig:fig1_processing}b-\ref{fig:fig1_processing}d). As \textit{in situ} process, the method also prevents unfavourable oxidation from ambient \ce{O2}\cite{May_InP_H2O_2014}. Under open-circuit (OC) conditions, the sample was illuminated with white light, initiating an oscillation (see Fig.~\ref{fig:fig1_processing}d) of the observed potential. Oscillation periods were in the range of 20\,s with amplitudes in the order of 100\,mV. After typically 5-6 oscillations, the OC potential reached its maximum. We attribute this phenomenon to a slow layer-by-layer etching, where surfaces with superior electronic quality are simultaneously prepared. The process shows similarity to the current-oscillations observed at the Si--electrolyte contact\cite{Aggour_PES_Si_oscillations_1995, Grzanna_Si_oscillations_2000}: In  this model, the anodic and the cathodic currents equilibrate at open circuit; the anodic partial current leads to oxide formation in competition with the cathodic processes of \ce{H2} evolution, excess minority carrier recombination, and the reduction of formed oxides. Oscillations occur if, within one cycle, the locally formed initial oxide islands are reduced more slowly, hence are more stable than the later formed ones, which exhibit more strain-induced defects. This constitutes a feedback mechanism that synchronises the oxidation and the cathodic etching process. This hypothesis is supported by the finding, that at a maximum of OCP, the thickness of the oxide is reduced to 0.4\,nm, in comparison to 0.8\,nm at the OCP minimum (see Supplementary Fig.~1). Analogous formation of intermediate islands, mediated by surface defects, has also been observed for the removal of oxides from III-V surfaces in gas-phase experiments\cite{Moeller_GaSb_deoxidation_2005}. The periodic change of the photovoltage is attributed to periodic thickness variations, corresponding potential drops across the interfacial film, and periodically increased recombination within a cycle when more defect-rich films are formed. The photovoltage is highest for the least oxidised surface.

The photovoltage oscillation step was directly followed by electrodeposition of the Rh electrocatalyst. The deposition was performed by applying potential pulses of 50\,ms duration over a time of typically 30\,s in the dark (see Methods). Subsequently, photoelectrodeposition of Rh was applied by periodically varied illumination between low and high light intensity, followed by 30\,s in the dark (see also Supplementary Fig.~2). The oscillating behaviour of the OCP upon illumination during the functionalisation is not a unique feature of the Rh-containing solution, but can also be induced in pure \ce{HClO4}. From a device perspective, however, the beneficial effect of the oscillation on the surface is maximised if the catalyst deposition is conducted directly thereafter, i.e. in the same electrolyte.

The process resulted in a composite structure of seed catalyst nanoparticles and a thin Rh covering layer (Figs.~\ref{fig:fig1_processing}c,e). XPS revealed that the initial surface functionalisation before Rh deposition led to a reduction of the In content at the surface and, preferentially, of the relative In oxide contribution (Fig.~\ref{fig:XPS}a). While the Al 2s line (Fig.~\ref{fig:XPS}b) is slightly reduced and develops a small shoulder towards higher binding energies, the P 2p signal shows a partial transformation into a mixture of InPO$_4$ and In(PO$_3$)$_3$, indicated by the peak between 133 and 134\,eV binding energy\cite{Chen_InP_oxidation_RAS_2002, PEC_book_chapter_2013}. Peak analysis shows that the relative \ce{PO_x} contribution is increased from initially 6\% to 67\% after functionalisation. 

Photoemission lines of the functionalised surfaces are, compared to those of the etched samples, shifted by $(0.75\pm0.1)$\,eV to higher binding energies (see Fig.~\ref{fig:XPS}). This shift originates in part from the higher n-doping of the functionalised, phosphate- and phosphite-rich surface, also observed for InP photocathodes\cite{PEC_book_chapter_2013}.

For the oxidised sample directly after etching, the Fermi level at the topmost surface is energetically lowered in comparison to the highly n-doped AlInP bulk due to a lower doping and a higher electron affinity of the oxide \cite{Vurgaftman_band_parameters_III-V_2001, Munoz_conditioning_InP_2013}. The completely functionalised surface that was transformed towards \ce{PO_x}, on the other hand, exhibits a higher n-doping and the Fermi level is shifted towards the vacuum level (cf. Fig.~\ref{fig:XPS} and Supplementary Fig.~3), improving the band alignment to the AlInP. The beneficial effect of the \ce{PO_x} phase is further supported by the observation that the signal strength of this species (see Fig.~\ref{fig:XPS}b) is directly correlated with the overall efficiency. In a quantitative analysis of the P 2p lines, the average thickness of the phosphate and phosphite layer was estimated to be $(1.3\pm0.2)$\,nm.

An analysis of transient photovoltages (Supplementary Fig.~4) shows an overshoot for not fully transformed surfaces, which indicates a trapping of charge carriers in surface states \cite{Li_surface_recombination_sc_electrodes_1984}. Together with the finding, that open-circuit voltages increase significantly for devices with \ce{PO_x} termination (cf. Supplementary Fig.~5), we conclude that the functionalisation layer also provides an electronic surface passivation. This transformation is a feature of the \textit{in situ }functionalisation prior to catalyst deposition, which avoids the formation of charge carrier recombination centres at strained In/Al--O--In/Al bonds, that would form upon exposure to atmospheric \ce{O2}\cite{Wood_surface_chemistry_water_InP_GaP_2014, May_InP_H2O_2014}, and turned out to be a crucial prerequisite for high efficiency. 

\subsection*{Cell performance}

\begin{figure}
\includegraphics[width=.6\textwidth]{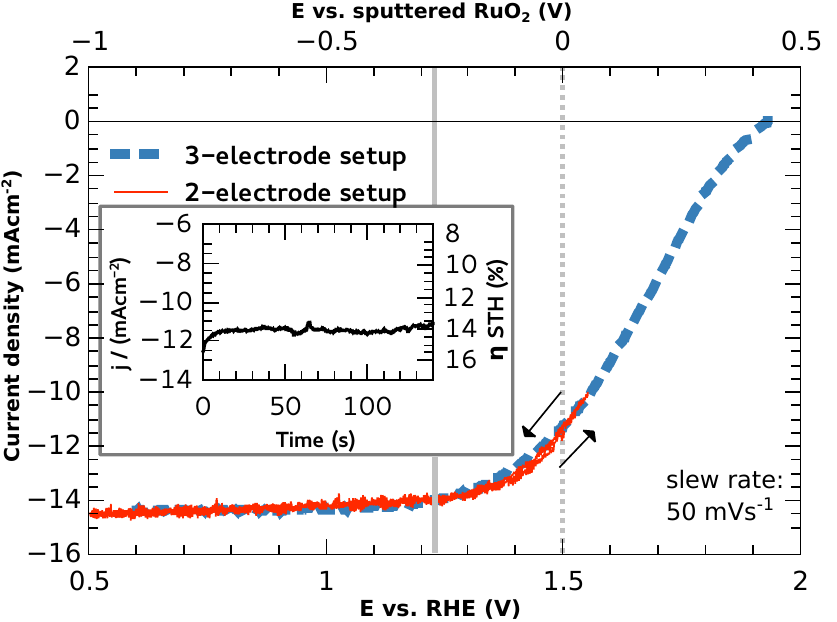}
\caption{\textbf{Output power characteristic.} The \textit{in situ}-modified two-junction tandem structures were evaluated in 1\,M HClO$_4$ under AM 1.5G illumination. The red curve shows a two-electrode configuration with a RuO$_2$ counter-electrode. The blue curve represents a cyclic voltammogram in a three-electrode setup with Pt counter-electrode plotted vs. the reversible hydrogen electrode redox potential (RHE). The grey, solid line marks +1.23\,V vs. RHE, the dashed line 0\,V vs. \ce{RuO2}. The curves intersect at 0\,V vs. \ce{RuO2} and 1.5\,V vs. RHE. The inset depicts the current over time for the unbiased two-electrode configuration.}
\label{fig:output}
\end{figure}

The output characteristics of the resulting devices (stacking and energy schematic sketched in Figs.~\ref{fig:fig1_processing} and \ref{fig:device_structure}) were evaluated in 1\,M HClO$_4$ under simulated sunlight (AM 1.5G, light intensity $I=100$\,mWcm$^{-2}$) (Fig.~\ref{fig:output}). Both, results of a two-electrode and a three-electrode potentiostatic measurement are shown. For the former, a current density of $j=11.5$\,mAcm$^{-2}$ without bias is reached, as seen in the inset. With the definition of the solar-to-hydrogen efficiency, $\eta = (j_{H_2}\times1.23\,V)/I$, this translates into an efficiency of $\eta=14\%$ under the assumption of 100\% Faradaic efficiency. A saturation current density beyond 14\,mAcm$^{-2}$, measured in a three-electrode setup, is close to its maximum value, indicating that light transmission through the catalyst is not a major issue. The open-circuit potential of 1.9\,V is, however, 200\,mV below the expected value, which is mainly ascribed to interfacial charge carrier recombination. In conjunction with the linear-exponential catalyst characteristic, this results in the S-shaped onset of the photocurrent\cite{Shaner_cv_coupled_photodiode_electrocatalyst_2013}. The difference between two- and three-electrode setup measurements is attributed to the overpotential of the oxygen evolution reaction (OER) as well as the solution resistance. The observation of 14\,mAcm$^{-2}$ at +1.23\,V vs. RHE results in an assisted photocathode efficiency of $>17\%$ for the three-electrode arrangement (blue curve in Fig.~\ref{fig:output}), demonstrating the potential efficiency of the device. An example for an incomplete functionalisation towards \ce{PO_x} is given in the Supplementary Fig.~5, showing a significantly reduced fill factor. A successful surface transformation combined, however, with catalyst overloading results in reduced optical transmission and an associated reduction of the photocurrent, while the open-circuit voltage is increased. 8\% of the samples, that had an active surface area between 30 and 100\,mm$^2$, showed highly successful functionalisation with photocurrent densities of 11\,mAcm$^{-2}$ and beyond, 25\% exceeded 10\,mAcm$^{-2}$ (see also Methods section).

\begin{figure}
\includegraphics[width=.55\textwidth]{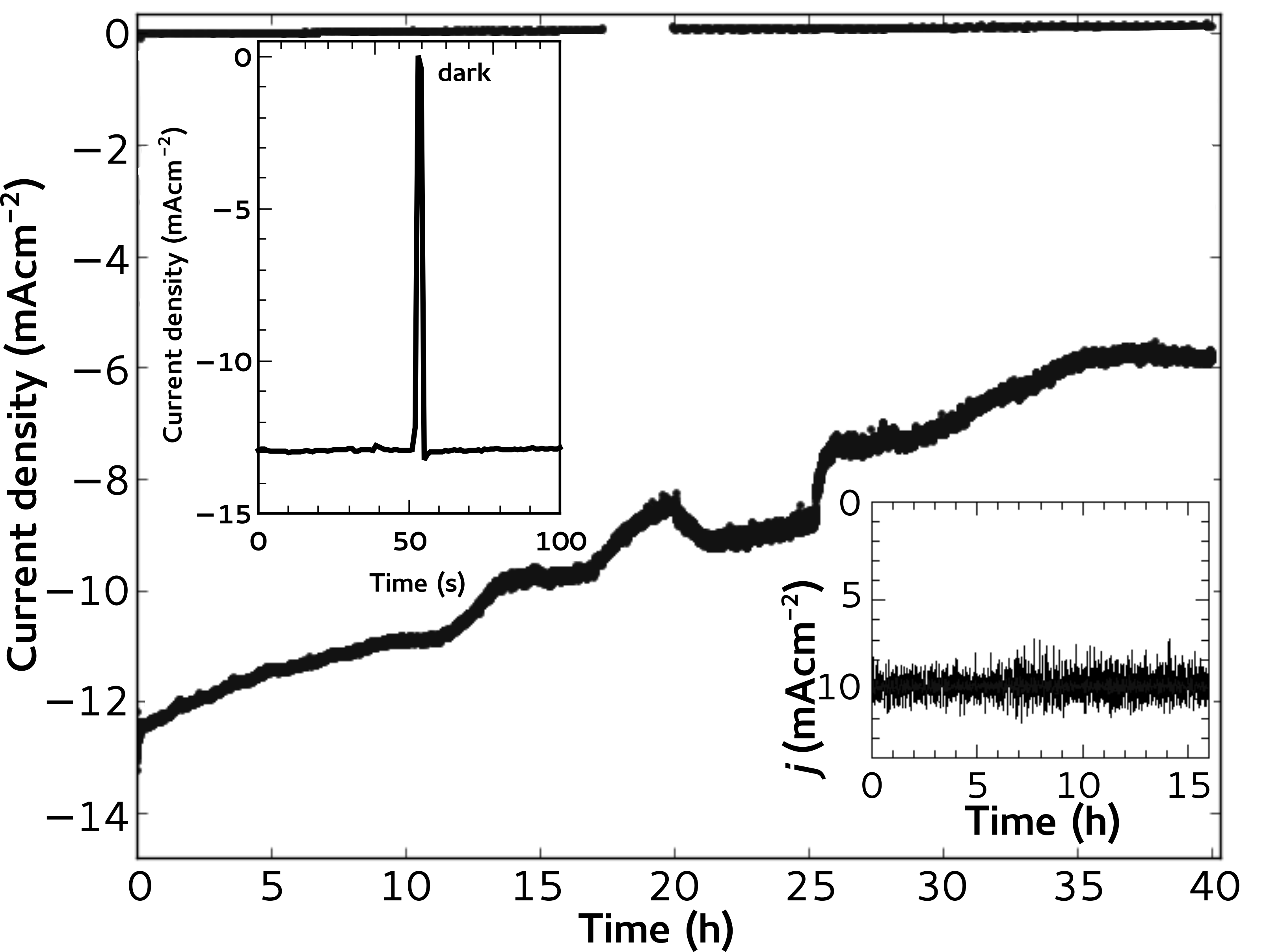}
\caption{\textbf{Stability assessment.} The AM 1.5G illumination in a vertical electrode geometry was chopped with 5\,s dark time and 200\,s exposure. Left inset: Zoom into the first light-dark cycle. Right inset: Current evolution in a horizontal, non-chopped three-electrode configuration. Potential +600\,mV vs. RHE.}
\label{fig:longterm}
\end{figure}
 
Chronoamperometric data over 40\,h are displayed in Fig.~\ref{fig:longterm}. For a vertical configuration, the photocurrent density degrades continuously, saturating at 0.5 of its original value after 35\,h of operation. Optical inspection of the electrode surface shows that steps in the decrease of the current are associated with severe detachment of catalysts from the surface, likely a result of chemical etching and stress from \ce{H2} bubbles propagating parallel to the surface, as already proposed in Ref.\cite{Khaselev1998}. In a horizontal setup (with slight catalyst overloading as indicated by only $j=10$\,mAcm$^{-2}$), no degradation of the photocurrent for over 16\,h of operation is noted (see inset of Fig.~\ref{fig:output}).

These data show the hitherto best stability reported for a monolithically integrated, high efficiency, fully immersed device. Longer stability, obtained with half cells\cite{Lewerenz_EESci_2010, Hu_leaky_TiO2_2014} cannot directly be compared with those of tandem structures because the absorbers used in those experiments were high quality single-crystalline Si, GaAs, InP, and GaP. In heteroepitaxy for tandem cells, multi-stack layer growth results in growth-induced defects that can protrude to the surface. Such sites can act as nucleation centres for subsequent defects, which develop during surface transformation, electrodeposition, and finally operation. Resulting pinhole formation leads to a reduced mechanical stability of the Rh catalyst due to undercutting at sites, where the corrosion protection layer has been incompletely formed. The mechanical detachment of the catalyst is more severe in a vertical configuration of the electrode, where gas bubbles, which have been observed to move directly along the surface, induce optical and mechanical interaction in particular at defect sites and further local corrosion. Our results point to the necessity to prepare custom-designed tandem absorber structures because the difference in surface chemistry at defect sites will also limit protection possibilities via atomic layer deposition. A possible solution could be the deposition of thin layers of earth-abundant materials by physical vapour deposition followed by underpotential deposition\cite{Kolb_underpotential_deposition_1974} of the catalyst, which would also reduce the overall noble-metal consumption.

\begin{figure}
\includegraphics[width=.5\textwidth]{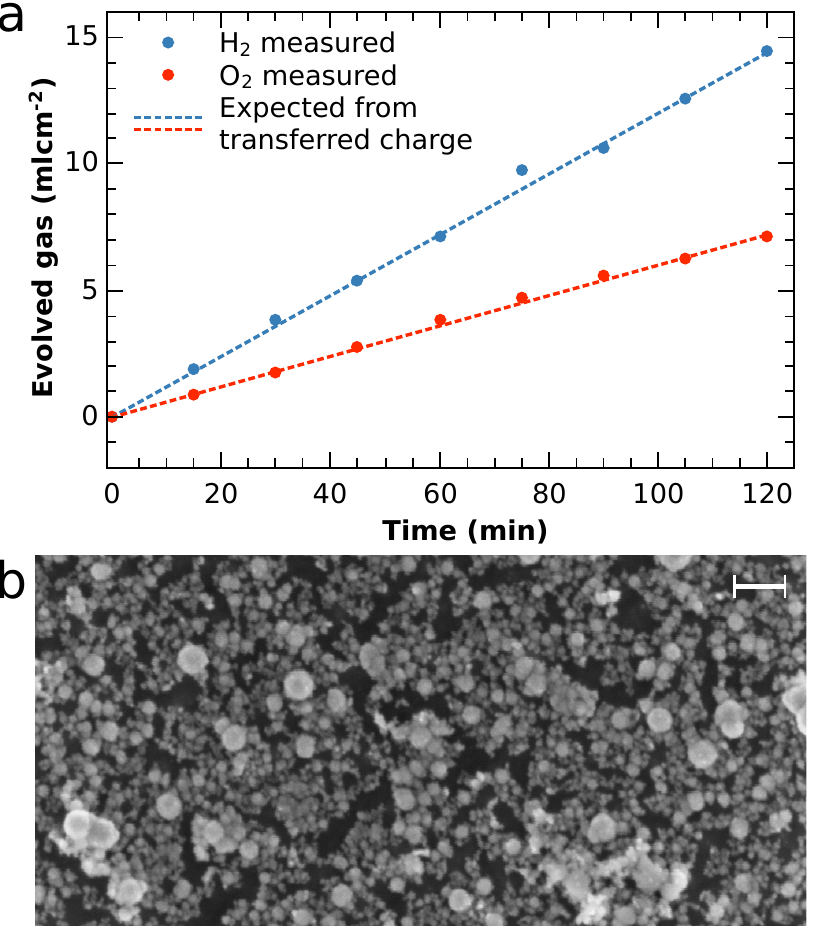} 
\caption{\textbf{Gas evolution and surface morphology.} (a) Gas evolution of the illuminated device under galvanostatic conditions. (b) Rh catalyst nanoparticles at the surface after the passage of -40\,Ccm$^{-2}$ under operation by scanning electron microscopy (backscattering image). Black areas show the underlying absorber, grey areas the catalyst nanoparticles, the scale bar indicates 200\,nm.}
\label{fig:gas_sem}
\end{figure}

High Faradaic solar-to-hydrogen efficiency is confirmed by the measurement of gas evolution (Fig.~\ref{fig:gas_sem}a), exhibiting a 2:1 ratio of \ce{H2} to \ce{O2} and an equivalent of the evolved gases to the overall charge transferred within the measurement error of ca. 5\%.

\section*{Discussion}

To estimate the Faradaic efficiency, we consider a potential lateral removal of the flat, epitaxial absorber structure (0.8$\mu$m for the top cell). In a cathodic corrosion process, where \ce{PH3} is formed, three electrons are liberated per Al/In atom\cite{Khaselev_stability_1998}, which is equivalent to 12 electrons per crystal unit cell. With a lattice constant $a=0.57$\,nm, this translates into a charge of 1.04\,\ce{mCcm^{-2}} for the corrosion of 1\,nm. Assuming a (conservative) average current density of 7\,\ce{mAcm^{-2}} during the long term experiment in Fig.~\ref{fig:longterm} over 40\,h and homogeneous corrosion, this would result in a removal of 1\,mm, if all the current would be due to corrosion. As the device still splits water after 40\,h, the corroded thickness is certainly less than 0.8\,$\mu$m, the thickness of the top cell. Hence, the lower boundary of the Faradaic efficiency for \ce{H2} evolution can be estimated to be $>99.9\%$. Gas measurements (Fig.~\ref{fig:gas_sem}a) confirm such a high Faradaic efficiency with the ratio of cathodic to anodic products of 2:1 and the overall amount of gases very close to the expected value derived from the integration of the current.

A \textit{general motif} of a successful functionalisation of InP-based surfaces appears to be the formation of phosphate and phosphite species, simultaneously providing stability as well as efficient charge-transfer\cite{Lewerenz_EESci_2010, PEC_book_chapter_2013, May_InP_H2O_2014}. For the AlInP surface studied here, the removal of mixed Al-/In-oxides and the concurrent incorporation of oxygen in the form of \ce{PO_x} on the first 1-2\,nm is a crucial prerequisite for stability and efficiency. While In--O--In bonds were indeed predicted to form potential charge carrier recombination centres\cite{Wood_surface_chemistry_water_InP_GaP_2014}, the phosphate species seem to provide a similar surface passivation as it was observed in the case of cobalt phosphate on metal-oxide electrodes\cite{Zhong_surface_passivation_Co-Pi_2011}.

For an energetic consideration of the interface formation during functionalisation, two junctions have to be regarded, the buried solid--solid interface between the pure, as-grown, AlInP window layer and its oxygen-containing surface layer, as well as the solid--liquid interface between this oxygen-modified surface layer and the electrolyte. In the case of the buried junction, the electron affinities, $\chi$, of the AlInP and the oxide/\ce{PO_x} species have to be considered. While AlInP exhibits $\chi=3.8$\,eV\cite{Vurgaftman_band_parameters_III-V_2001}, it changes to $\chi\approx4.4$\,eV for indium oxide\cite{Munoz_conditioning_InP_2013}. Such differences in electron affinity between III-V semiconductors and their oxides with their potential impact on the photovoltage have already been addressed in the literature\cite{Kaiser_GaP_2012}. The transformation of the surface layer towards a \ce{PO_x}-rich phase with its reduced electron affinity \cite{Robach_InPOx_properties_1992} does, however, shift the conduction band back towards the vacuum level, reducing the band offset at the buried junction. This is supported by XPS, showing a binding energy of the In 3d peak of 443.8, 444.2 and 444.6\,eV for the freshly etched, the partly functionalised and the fully functionalised surface, respectively. This trend is confirmed by Mott-Schottky analysis, where the flat band of -0.7\,V vs. RHE for the fully functionalised surface corresponds to 3.9\,eV vs. the vacuum level, which is very close to what is expected for n-doped AlInP.

Due to the high n-doping in the AlInP and the ultrathin ($<2$\,nm) oxide/\ce{PO_x} layer, the junction between the absorber surface and the Rh nanoparticles is mainly defined by the Fermi level of the semiconductor\cite{Calvet_Si-based_tandems_as_photocathodes_2014}, and space charge regions are expected to be thin enough for tunneling. Consequently, the reduction of the surface recombination velocity by the \ce{PO_x}-rich phase is the main beneficial effect of the surface functionalisation.

Our findings suggest that the functionalisation should be transferable to other PV-based III-V (tandem) absorbers, which exhibit an InP-containing top layer or an In-rich compound surface. In cases where the absorber does initially not exhibit such a top layer, it is possible to terminate absorber growth with a thin, not necessarily lattice-matched, InP layer on top. As the functionalisation mainly depends on the surface chemistry and the Fermi level at the surface (which can be tuned by doping), the InP layer could serve as a sacrificial functionalisation precursor. Such an approach would consequently widen the material choice for the underlying tandem structure.

With the near-optimum, low light-attenuation catalyst loading found here, the amount of Rh for 1\,MW electrochemical power output would be in the order of 1\,kg. This mass could be considerably reduced by the use of core-shell catalyst nanoparticles prepared, for example, by under-potential deposition and galvanic exchange with a core of an earth-abundant material\cite{Zhong_core-shell_nanoparticles_2001}. Further cost reductions from the substrate side could be achieved by the employment of lift-off techniques or the switch to Si substrates, targeting in combination with light concentration the high-efficiency route to low-cost \ce{H2}\cite{PEC_book_chapter_2013, Pinaud_hydrogen_production_facilities_2013}.

The results demonstrate that \textit{in situ} modification of surfaces enables exceptionally high efficiency for unassisted solar water splitting. The combination of (photo)electrochemical functionalisation with surface science analyses provides atomic level insights and feedback for the optimisation of the charge transfer processes and interfaces. A further improvement of the functionalisation has the potential to achieve even higher stabilities without the need for protective layers by atomic layer deposition, which would introduce additional processing. If the gap between potential (17.4\%) and achieved (14\%) solar-to-hydrogen can be further reduced, devices producing solar \ce{H2} below 4\$/kg\cite{Pinaud_hydrogen_production_facilities_2013} could become feasible.

\section*{Methods}

\subsection*{Epitaxial absorber growth and ohmic contact}

The tandem absorber was grown epitaxially by metal-organic vapour phase epitaxy on a Ge substrate. The lattice-mismatch to the substrate was accounted for by an intermediate step-grading buffer, gradually changing the lattice constant from the substrate to the absorber. For a detailed description, the reader is referred to Ref.\cite{Dimroth_metamorphic_cell_2001}. The ohmic back-contact was prepared by evaporating Ni and Au on the substrate, followed by annealing.

\subsection*{Etching and chemical oxidation}

Prior to etching, samples were cleaned in 2-propanol and water to remove initial surface contamination. The etching procedure removed the cap layer stopping on the AlInP window layer. The etching solution was composed of NH$_4$OH (25\%, Sigma Aldrich p.a. grade), H$_2$O$_2$ (30\%, Merck p.a. grade) and H$_2$O in a ratio of 1:1:10 and prepared directly before the etching. All the H$_2$O was ultrapure 'Milli-Q' grade. Immediately after the removal of the window layer, samples were rinsed in H$_2$O, saturated with \ce{O2} by purging. The resulting hydrophilic surface was then slowly dried in a stream of \ce{O2} for several minutes. Surface transformation was performed in a three-electrode-setup\cite{Bard_and_Faulkner_book_2001} consisting of a Pt counter-electrode and an Ag/AgCl reference electrode in an aqueous RhCl$_3$ solution. Samples were encapsulated in a two-component epoxy resin or, if XPS analysis was performed, clamped against a sealing ring.

\subsection*{Surface transformation and Rh deposition}
The electrolyte for functionalisation and Rh deposition was an aqueous solution of 5\,mM Rh(III) chloride trihydrate (99.98\%, Sigma Aldrich) + 0.5\,M KCl (99.5\%, Alfa Aesar) + 0.5 vol\% 2-Propanol (Sigma Aldrich p.a. grade). White light was provided by a tungsten iodide lamp. In the three-electrode setup, the voltage of the working electrode (the sample) is controlled via a potentiostat (PAR Versastat 3) against a reference electrode, in this case Ag/AgCl (+0.197\,V vs. the normal hydrogen electrode)\cite{Bard_and_Faulkner_book_2001}. The current then flows between working electrode and a Pt counter-electrode immersed in the same solution. Samples were typically embedded into a black two-component epoxy resin. In the cases where XPS analysis was performed, the samples were not embedded into epoxy resin to avoid outgassing and clamped against a sealing O-ring in the PTFE beaker as the cell compartment.

The surface functionalisation was initiated by immersing the sample in the RhCl$_3$ solution under open-circuit conditions in the dark. Illumination with white light caused a shift of the OCP to more positive potentials, modulated by an oscillation. When the OC potential exhibited a maximum, illumination was turned off terminating the photochemical optimisation (Fig.~\ref{fig:fig1_processing}d). XPS analysis after this step did not reveal any Rh ($<5\%$ of a monolayer). Further (photo-)electrochemical surface conditioning and catalyst deposition were initiated by the deposition of an Rh catalyst seed layer in the dark, applying 50\,ms potential pulses at -0.3\,V vs. SCE (separated by 1\,s at -0.1\,V vs. SCE) for several minutes (see Supplementary Fig.~2). Maintaining the potential pulsing, illumination (also pulsed) was turned on again, leading to an accelerated deposition of Rh in the form of nanoparticles with an average size of ca. 20\,nm (see Fig.~\ref{fig:gas_sem}b). The procedure was completed by another pulsed deposition in the dark.

As the reduction of Rh$^{3+}$ to metallic Rh competes with the \ce{H2} evolution reaction, the Faradaic efficiency for the deposition is not unity. This explains the transferred charges in the order of 50\,mCcm$^{-1}$, which would otherwise result in a (closed) Rh layer with a thickness in the order of 10\,nm. The optimum catalyst loading exhibits a relatively narrow process window: A larger loading results in a higher open-circuit voltage and a steeper onset of the photocurrent, but this increase is over-compensated by a lower photocurrent due to increased light attenuation by the catalyst.

\subsection*{Output characteristic}
The setup for determination of the output characteristics was again a three-electrode setup with Pt as counter-electrode and Ag/AgCl as reference electrode or a two-electrode setup with a \ce{RuO2} counter-electrode (1\,cm$^2$). A commercial Wacom WXS-50S solar simulator provided the AM 1.5G spectrum at the location of the sample, which was calibrated by quantitatively matching the measured spectrum, which was acquired with an Ocean Optics USB2000+RAD spectrometer, to AM 1.5G. The PTFE cell (as described in Ref.~\cite{Krol_PEC_hydrogen_production_2011}) was equipped with a quartz plate as front window and an O-ring of 6\,mm diameter defining the active surface. Electrical connection was assured by a gold finger pressed against the back of the ohmic contact of the solar cell. In the cases, where samples were embedded in epoxy resin, an all-quartz cell was employed. The electrolyte was 1\,M HClO$_4$ and, in most cases for the vertical configuration, 1\,mM Triton X 100 as a surfactant to reduce the bubble size. Long-term stability in the horizontal setup was evaluated without the use of a surfactant. 18\,h into the chopped long term experiment in Fig.~\ref{fig:longterm}, an intermediate failure of the chopper lead to constant illumination for 2\,h.

Statistics: Out of 65 analysed samples, 25\% showed current densities beyond 10\,mAcm$^{-2}$ and 8\% above 11\,mAcm$^{-2}$.

Gas evolution at the electrodes was quantified by a setup employing two pipettes and manual pressure compensation, after King and Bard\cite{King_gas_volume_measurement_1964}. The cell, containing 1\,M \ce{HClO4}, was first purged with Ar. Afterwards, the solution was saturated with \ce{H2} and \ce{O2} by dark electrolysis, using the \ce{RuO2} counter-electrode and an Ir working electrode. Subsequently, the working electrode was switched to the tandem structure, which was then illuminated. The gas volume was measured under ambient conditions.

\subsection*{Photoelectron spectroscopy}

For photoelectron spectroscopy, a monochromated Al K$_\alpha$ source was used. Due to the high photoactivity of the samples inducing a surface photovoltage, energy calibration was achieved by setting the binding energy (relative to the Fermi level E$_F$) of the C 1s line to 284.8\,eV. For the peak analysis, Voigt profiles in combination with a linear background were employed.

For the XPS measurements at different stages of the OCP oscillation (Supplementary Fig.~1), the electrochemical experiment was abruptly interrupted and the sample transferred to UHV. To rule out a potential build-up of oxide by subsequent interruptions of the electrochemical experiment, the order of the XPS measurement with respect to maximum/minimum was reversed for different samples, confirming the results.


\section*{Acknowledgements}
The authors are grateful for support by R. van de Krol and for experimental support by M. Kernbach and H. Kriegel. M. Niemeyer, C. Karcher, and J. Ohlmann supported the solar cell development. K. Harbauer and T. M\"unchenberg assisted with ohmic contact preparation. U. Bloeck performed the TEM measurements, S. Brunken provided the \ce{RuO2} counter-electrode. The authors also thank P. Bogdanoff, K. Fountaine, and C. McCrory for helpful discussions. M.M.M. acknowledges a scholarship by the Studienstiftung des deutschen Volkes, H.J.L. acknowledges support by the DFG (project Nr. 1192-3/4). The discussion and interpretation of the data, feedback with experimentation as well as article layout and writing was also supported by the Joint Center for Artificial Photosynthesis, a DOE Energy Innovation Hub, supported through the Office of Science of the U.S. Department of Energy under Award Number DE-SC0004993.

\section*{Author Contributions}
T.H., H.J.L., and M.M.M. designed the study. M.M.M. executed the experiments and did the data analysis. D.L. and F.D. prepared the tandem absorber. M.M.M. and H.J.L. wrote the paper and all authors commented on the manuscript.

\section*{Competing Financial Interests} 

The authors declare no competing financial interests.\\
\newpage


\section*{Supplementary material}

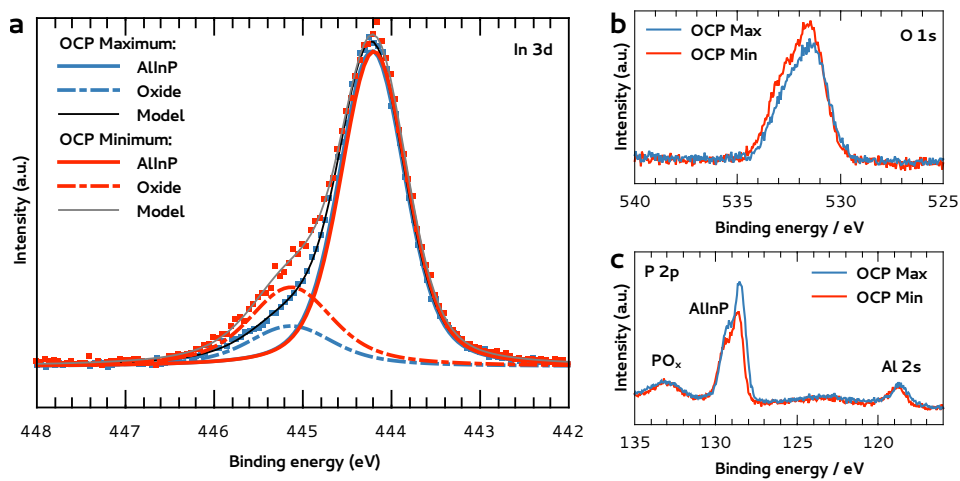

Supplementary Figure 1: **XPS at different stages of the OCP oscillation.** (a) The blue (red) curve shows the In 3d core level at an OCP maximum (minimum). The dashed lines indicate the oxide contribution. (b) O 1s core level. (c) P 2p and Al 2s core levels. To evaluate the oxide thickness, we used the model of a homogeneous overlayer of thickness $d$ [1], assuming the same photoelectron cross sections for oxide and bulk. The electron attenuation length, $\lambda$, was calculated to be $2.5\,\text{nm}$ [2], and the intensities of oxide and bulk are denoted $I_o$ and $I_b$, respectively: $d = \lambda \ln\left(\frac{I_o}{I_b} + 1\right)$. For the spectra displayed above, this results in an overlayer thickness of ca. $(0.4 \pm 0.1)\,\text{nm}$ for the OCP maximum and $(0.8 \pm 0.1)\,\text{nm}$ for the minimum.

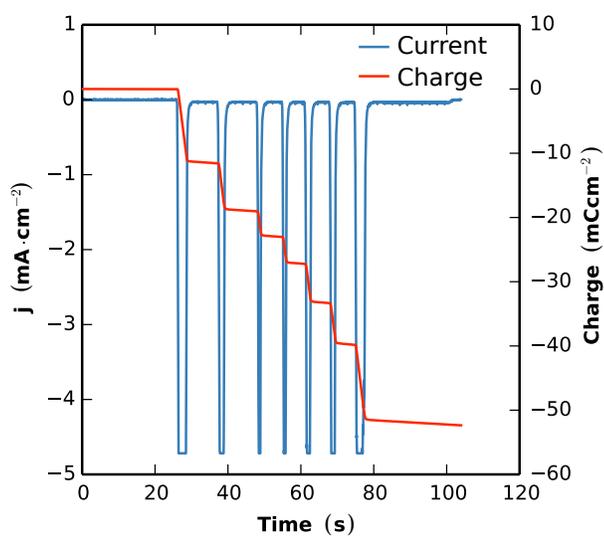

Supplementary Figure 2: **Photoelectrochemical catalyst deposition.** The blue curve shows the current density, the red curve the integrated charge.

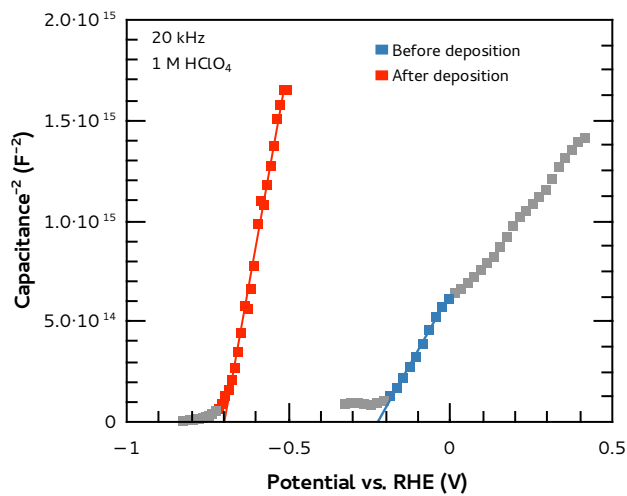

Supplementary Figure 3: **Evaluation of the flat-band potential.** The Mott-Schottky plots were measured in the dark after short OCP treatment (blue) and after Rh deposition (red). Linear fits show flat-band potentials of -0.22 V and -0.70 V vs. RHE before and after full surface transformation, respectively. Gray indicates data points that were not considered in the fit.

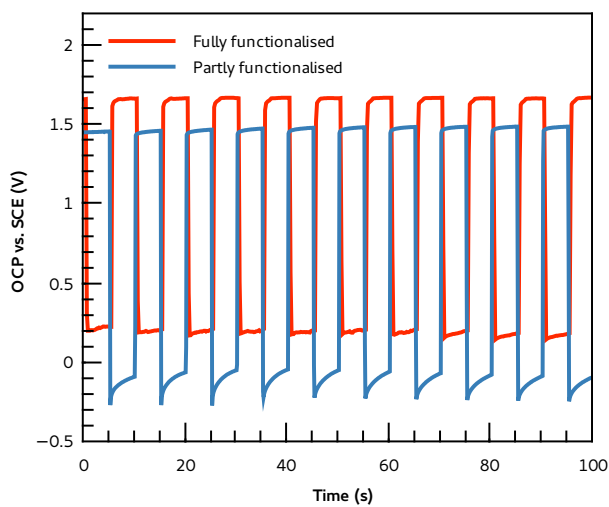

Supplementary Figure 4: **Open-circuit potential (OCP) transients.** The measurement employed a saturated calomel electrode (SCE) and chopped simulated AM 1.5G illumination for a partly (blue) and fully (red) functionalised sample. The overshoot of the partly functionalised sample indicates trapping of charge carriers in surface states [3]. The offset between both samples under blocked illumination arises due to very low residual illumination within the solar simulator setup, which was identical in both cases, demonstrating the higher photovoltage of the functionalised sample.

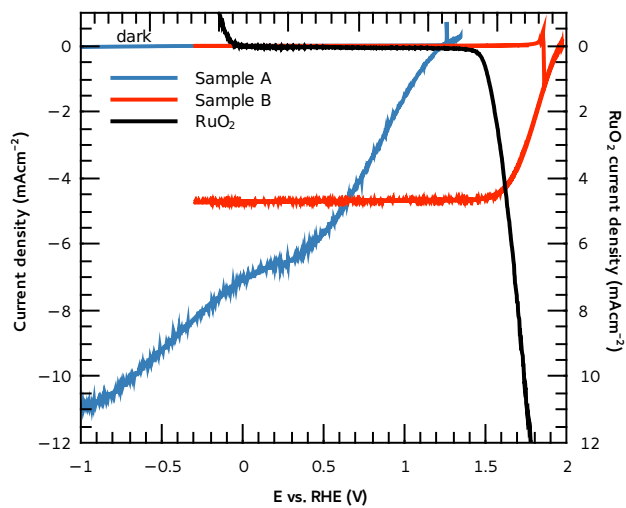

Supplementary Figure 5: **Comparison of differently functionalised samples.** Cyclic voltammograms under sim. sunlight for a sample without successful surface functionalisation (sample A, blue curve) reducing the fill factor and with catalyst overloading (sample B, red curve) reducing light transmission, but also recombination as evidenced by transient photocurrent analysis. The black curve shows the $RuO_2$ counter-electrode.

**Supplementary References**



\end{document}